\documentclass{pepsart}
\usepackage{amsthm,amsmath}
\usepackage[utf8]{inputenc} 
\usepackage{wasysym}
\usepackage{graphicx}
\usepackage{journals2}
\usepackage[nomarkers]{endfloat}

\startlocaldefs
\newcommand\rev[1]{{\color{black}#1}}
\newcommand\revb[1]{{\color{black}#1}}
\endlocaldefs

\begin{document}

\begin{frontmatter}

\begin{fmbox}
\dochead{Research}

\title{Effect of width, amplitude, and position of a core mantle boundary hot spot on core convection and dynamo action }


\author[
      addressref={aff1},                   
   email={w.dietrich@leeds.ac.uk}   
]{\inits{W}\fnm{Wieland} \snm{Dietrich}}
\author[
   addressref={aff2},
   email={wicht@mps.mpg.de}
]{\inits{J}\fnm{Johannes} \snm{Wicht}}
\author[
   addressref={aff1,aff3},
   email={amtkh@leeds.ac.uk}
]{\inits{K}\fnm{Kumiko} \snm{Hori}}

\address[id=aff1]{
  \orgname{Department of Applied Mathematics, University of Leeds}, 
  \street{Woodhouse Lane},                     %
  \postcode{LS9 2JT}                                
  \city{Leeds, West Yorkshire},                              
  \cny{United Kingdom}                                    
}
\address[id=aff2]{%
  \orgname{Max Planck Institute for Solar System Research},
  \street{Justus-von-Liebig-Weg 3},
  \postcode{37077}
  \city{G\"ottingen},
  \cny{Germany}
}

\address[id=aff3]{%
  \orgname{Solar-Terrestrial Environment Laboratory, Nagoya University},
  \street{Furo-cho, Chikusa-ku},
  \postcode{464-8601}
  \city{Nagoya},
  \cny{Japan}
}




\end{fmbox}

\begin{abstractbox}

\begin{abstract} 
  Within the fluid iron cores of terrestrial planets, convection and the resulting generation of global magnetic fields are controlled by the overlying rocky mantle. The thermal structure of the lower mantle determines how much heat is allowed to escape the core. Hot lower mantle features, such as the thermal footprint of a giant impact or hot mantle plumes, will locally reduce the heat flux through the core mantle boundary (CMB), thereby weakening core convection and affecting the magnetic field generation process. In this study, we numerically investigate how parametrised hot spots at the CMB with arbitrary sizes, amplitudes, and positions affect core convection and hence the dynamo. The effect of the heat flux anomaly is quantified by changes in global flow symmetry properties, such as the emergence of equatorial antisymmetric, axisymmetric (EAA) zonal flows. For purely hydrodynamic models, the \rev{EAA symmetry} scales almost linearly with the CMB amplitude and size, whereas self-consistent dynamo simulations typically \rev{reveal either suppressed or drastically enhanced EAA symmetry depending mainly on the horizontal extent of the heat flux anomaly}. Our results suggest that the length scale of the anomaly should be on the same order as the outer core radius to significantly affect flow and field symmetries. \rev{As an implication to Mars \revb{and in the range of our model}, the study concludes that an ancient core field modified by a CMB heat flux anomaly is not able to heterogeneously magnetise the crust to the present-day level of north--south asymmetry on Mars. The resulting magnetic fields obtained using our model either are not asymmetric enough or, when they are asymmetric enough, show rapid polarity inversions, which are incompatible with thick unidirectional magnetisation.} 
\end{abstract}



\end{abstractbox}
%

\end{frontmatter}


\section*{Background}
Within our solar system, the three terrestrial planets, Earth, Mercury, and Mars, harbour or \rev{once} harboured a dynamo process in the liquid part of their \rev{iron-rich} cores. Vigorous core convection shaped by rapid planetary rotation is responsible for the generation of global magnetic fields. Unlike the dynamo regions of giant planets or the convective zone of the sun, the \rev{amount of heat escaping the cores of terrestrial planets is determined by the convection of the overlying mantle}. As this vigorous core convection assures efficient mixing and hence a virtually homogeneous temperature $T_{core}$ \rev{at the core side of the core mantle boundary (CMB)}, the temperature gradient at and thus the flux through the CMB are entirely controlled by the lower mantle temperature $T_{lm}$. The heat flux through the CMB is then
\begin{equation}
q_{cmb} = k \frac{T_{lm}-T_{core}}{\delta_{cmb}} \ ,
\end{equation}
where $\delta_{cmb}$ is the vertical thickness of the thermal boundary layer on the mantle side and $k$ is the thermal conductivity.
 Hot mantle features, such as convective upwellings, thermal footprints of giant impacts, or chemical heterogeneities, locally reduce the heat flux through the CMB \rev{(e.g., Roberts and Zhong 2006, Roberts et al.\ 2009)}. Seismologic investigations of Earth have revealed strong anomalies in the lowermost mantle that can be interpreted in terms of temperature variations (Bloxham 2000\nocite{Bloxham2000}). In recent years, several authors have investigated the consequences of laterally varying the CMB heat flux. A seismologically based pattern has been used in many studies on the characteristics of Earth (e.g., Olson et al.\ 2010\nocite{Olson2010}, Takahashi et al.\ 2009\nocite{Takahashi2008}); this pattern shows a strong effect on the core convection and secular variation of the magnetic field (e.g., Davies et al.\ 2008\nocite{Davies2008}). For Mars, low-degree mantle convection or giant impacts may have significantly affected the core convection and the morphology of the induced magnetic field (Harder and Christensen 1996\nocite{Harder1996}, Roberts and Zhong 2006\nocite{Roberts2006}, Roberts et al.\ 2009\nocite{Roberts2009}). For example, the strong southern hemispherical preference of the crustal magnetisation can be explained by an ancient dynamo that operated more efficiently in the southern hemisphere. However, this remains a matter of debate because post-dynamo processes may simpl\rev{y} have reduced a once more homogeneous crustal magnetisation in the northern hemisphere (Nimmo et al.\ 2008\nocite{Nimmo2008}, Marinova et al.\ 2008\nocite{Marinova2008}). The asymmetry of Mercury's magnetic field, which is significantly stronger in the north than in the south, could also be partly explained by a non-homogeneous CMB heat flux (e.g., Cao et al.\ 2014\nocite{Cao2014}, Wicht and Heyner 2014\nocite{Wicht2014}).

The magnetic field generation process inside Earth's outer core relies on thermal and compositional convection. Thermal convection is driven by secular cooling or the latent heat released upon inner core freezing. Compositional convection arises because the light elements mixed into the outer core alloy are not as eas\rev{ily} contained in the inner core. A large fraction is thus released at the inner core front. For planets with no solid inner core, like \rev{ancient} Mars or early Earth, only the thermal component of \rev{the buoyant force} can power convection and hence the \revb{dynamo} process (e.g., Breuer et al. 2010\nocite{Breuer2010}). The buoyancy sources are then not concentrated at the bottom but homogeneously distributed over the core shell. \rev{When modelling these processes, secular cooling can be modelled using a buoyancy source equivalent to internal heating, whereas a basal heating source can be used when thermal and/or compositional buoyancy fluxes arise from the inner core boundary.} Kutzner and Christensen (2000)\nocite{Kutzner2000} and Hori et al.\ (2012)\nocite{Hori2012} investigated the dynamic consequences when a core model is driven by either internal or basal heating. In general, the effects of large-scale CMB heat flux anomalies on convection and magnetic field generation are stronger when the system is driven by internal (as in ancient Mars) rather than basal heating, which is more realistic for present-day Earth (Hori et al.\ 2014\nocite{Hori2014}). For example, the equatorial symmetry of the flow is more easily broken in the former than the latter case (Wicht and Heyner 2014).

The Mars Global Surveyor (MGS) mission revealed a remarkable equatorial asymmetry in the distribution of magnetised crust (Acu\~na et al.\ 1999\nocite{Acuna1999}, Langlais et al.\ 2004\nocite{Langlais2004}). \rev{Interestingly, the crustal topographic dichotomy is well aligned with the pattern of crustal magnetisation (Citron and Zhong 2012\nocite{Citron2012}). This is the magnetic imprint of an ancient core dynamo driven by thermal convection in the core or tidal dissipation (Arkani-Hamed 2009\nocite{Arkani-Hamed2009}), which ceased to exist and further magnetise the crust roughly 3.5 Gyrs ago (Lillis et al.\ 2008\nocite{Lillis2008}). Assuming the hemispherical crustal magnetisation is of internal origin,} most numerical models attempting to design a Mars core dynamo model quantified the success of their efforts by comparing the resulting modelled magnetic fields to the actual crustal magnetisation pattern (Stanley et al.\ 2008, Amit et al.\ 2011, Dietrich and Wicht 2013\nocite{Dietrich2013}, Monteux et al.\ 2015\nocite{Monteux2015}). Hereafter, the study by \rev{Dietrich and Wicht (2013)} is abbreviated as DW13. The CMB heat flux in such models is typically modified by a large-scale sinusoidal perturbation increasing the CMB heat flux in the southern hemisphere and reducing it in the northern hemisphere. Because one hemisphere is more efficiently cooled than the other, a strong latitudinal temperature anomaly arises and in turn drives fierce zonal flows via a thermal wind. Such equatorially antisymmetric, axisymmetric (EAA) flows are reported to reach up to 85\% of the total kinetic energy (Amit et al.\ 2011, DW13) in self-consistent dynamo models. \rev{Although such flows are indicative of the induction of hemispherical fields, it remains unclear to what extent their prevalence is due to the (probably unrealistic) large, strong, and axisymmetric forcing patterns.} As a consequence of this forcing, the induction process is more concentrated in the southern hemisphere, leading to a hemispherical magnetic field. Even though such hemispherical fields can match the degree of hemisphericity in the crustal magnetisation at the planetary surface, they also show a strong time variability \rev{(periodic oscillations)}. If the typical stable chron epoch is much smaller than the typical crustal build-up time, the system is not able to magnetise the crust to the required intensity (DW13), which requires the relatively homogeneous magnetisation of a layer with a thickness of at least 20\,km (Langlais et al.\ 2004). 

In this study, we focus on somewhat more complex but also more realistic CMB heat flux variations. The heat flux is reduced in a \rev{more} localised area of varying position and \rev{horizontal} extent. Such anomalies may more realistically reflect the effect of mantle plumes or impacts\rev{, but may not yield the fundamental equatorial asymmetry in the temperature as efficiently as the simplistic $Y_{10}$ pattern. Further tilting the anomalies away from the axis of rotation may result in the superposition of EAA and \revb{equatorially symmetric, non-axisymmetric} (ESN) temperature and flow patterns.} We also investigate the influence of the shell geometry and the vigour of convection. \rev{More generally, we aim to estimate how large, strong, and aligned CMB heat flux anomalies must be to affect core convection and the magnetic field process. It is also of interest to quantify the interaction of flow and field under the control of a laterally variable CMB heat flux.} In particular, we address the question of whether the conclusions of DW13 regarding the crustal magnetisation still hold when applied to Mars. Various \rev{comparable} models have been investigated over the last decade. Those focusing on exploring parameter dependencies, e.g., Amit et al.\ (2011) and DW13, mainly investigated magnetic cases with a CMB heat flux described by a fundamental spherical harmonic. Those studies featured tilted cases and various amplitudes. \rev{The recent study by Monteux et al.\ (2015)\nocite{Monteux2015} presented simulations featuring anomalies of a smaller horizontal length scale. However, a comprehensive parametrisation of the anomaly width, amplitude, and position has not yet been reported.} The rather dramatic results of DW13 may only hold when anomalies of a planetary scale are at work. We therefore also test the robustness of their results with respect to the most common model assumptions.

\section*{Method and Model}
We model the liquid outer core of a terrestrial planet as a spherical shell \revb{(with inner and outer radii of $r_{icb}$ and $r_{cmb}$, respectively)} containing a viscous, incompressible, and electrically conducting fluid. The core fluid is subject to rapid rotation, vigorous convection and Lorentz forces due to the induced magnetic fields. The evolution of the fluid flow is thus given by the dimensionless Navier--Stokes equation:
\begin{equation}
\mathrm{E} \left( \frac{\partial \vec u}{\partial t} + \vec u \cdot \vec \nabla \vec u \right)
= -\vec \nabla \Pi + E \nabla^2 \vec u - 2 \hat z \times \vec u + \frac{\mathrm{Ra E}}{\mathrm{Pr}} \frac {\vec r}{r_{cmb}} T
+ \frac 1 {\mathrm{Pm}} (\vec \nabla \times \vec B ) \times \vec B \ ,
\label{nseq}
\end{equation}
where $\vec u$ is the velocity field, $\Pi$ is the non-hydrostatic pressure, $\hat z$
is the direction of the rotation axis, $T$ is the super-adiabatic temperature fluctuation, and
$\vec B$ is the magnetic field.

The evolution of the thermal energy is affected by temperature diffusion and advection by the flow, such that
\begin{equation}
\label{heat}
\frac {\partial T}{\partial t} + \vec u \cdot \vec \nabla T= \frac 1 {\mathrm{Pr}} \nabla^2 T
+ \epsilon \ ,
\end{equation}
where $\epsilon$ is a uniform heat source density. 
The generation of magnetic fields is controlled by the induction equation
\begin{equation}
\frac {\partial \vec B}{\partial t} = \vec \nabla \times \left( \vec u \times \vec B \right)
+ \frac 1 {\mathrm{Pm}} \nabla^2 \vec B \ .
\label{ind_eq}
\end{equation}

We use the shell thickness $D=r_{cmb}-r_{icb}$ as the length scale, the
viscous diffusion time $D^2/ \nu$ as the time scale, and $(\rho \mu \lambda \Omega)^{1/2}$
as the magnetic scale. The mean super-adiabatic CMB heat flux density $q_0$ serves
to define the temperature scale $q_0 D / c_p\rho\kappa$. Furthermore, $\nu$ is the viscous diffusivity,
$\rho$ is the constant background density, $\mu$ is the magnetic permeability, $\lambda$ is the
magnetic diffusivity, $\Omega$ is the rotation rate, $\kappa$ is the thermal diffusivity,
and $c_p$ is the heat capacity.

Non-penetrative and no-slip velocity boundary conditions are used, and the magnetic fields are matched to the potential field outside the fluid region. For an internally heated Mars-like set-up, we fix the heat flux at both boundaries and power the system exclusively by internal heat sources. Because the flux at the inner boundary is set to zero, this leads to a simple balance of heat between the source density $\epsilon$ and the mean CMB heat flux $q_{0}$. To model the secular core cooling, we fix the mean heat flux at the outer boundary ($q_0=1$) and balance the heat source such that
\begin{equation}
 \epsilon= \frac{1-\beta}{1-\beta^3} \frac{q_0}{3 \mathrm{Pr}} \ ,
\end{equation}
where \rev{$\beta=r_{icb}/r_{cmb}$} is the aspect ratio of the spherical shell.

The non-dimensional control parameters are the Prandtl number $\mathrm{Pr}=\nu/\kappa$, which is the ratio between the viscous and thermal diffusivities, and the magnetic Prandtl number $\mathrm{Pm}=\nu/\lambda$, which is the ratio of the viscous and magnetic diffusivities. The Ekman number $\mathrm{E}=\nu / \Omega D^2$ relates the viscous and rotational time scales, and the Rayleigh number $\mathrm{Ra}= \alpha g q_o D^4 / \nu \kappa^2$ controls the vigour of convection. We fix $\mathrm{Pr}=1$ and $mathrm{E}=10^{-4}$ and use $mathrm{Pm}=2$ for the dynamo cases. The Rayleigh number is varied between $mathrm{Ra}=2 \times 10^7$ and $1.6 \times 10^8$.

\subsection*{Modelling the Anomaly}
\begin{figure}[ht!]
\includegraphics[width=0.8\textwidth]{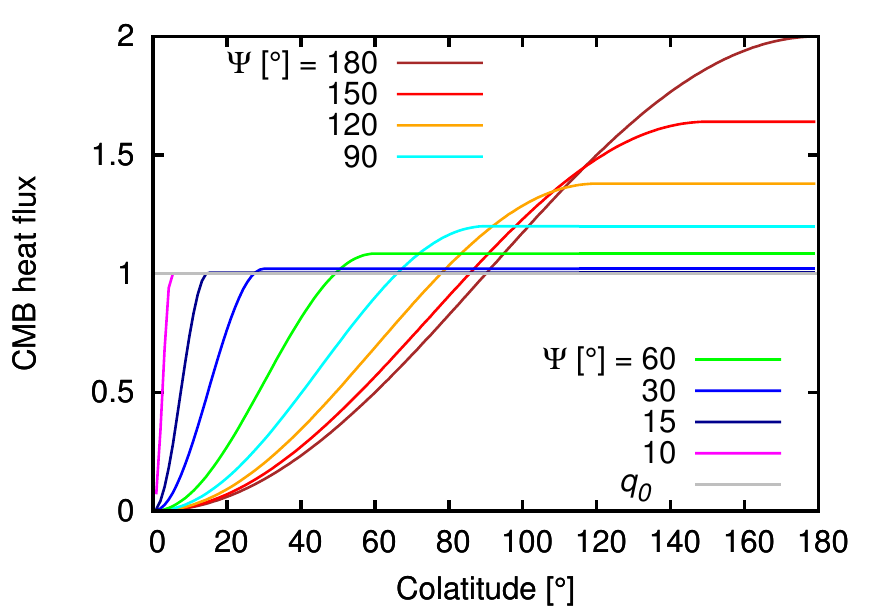}
  \caption{\csentence{Latitudinal profile of axisymmetric CMB heat flux}. The lines show the total CMB heat flux modified by an anomaly with different opening angles $\Psi$ (each value of $\Psi$ is represented by a different colour). Note that all profiles are normalised such that there is no net contribution from the anomaly to the mean CMB heat flux ($q_0=1$, grey). Parameters: $\tau=0$, $q^\ast=1$.}
\label{figqprof}
\end{figure}

In recent studies focusing on the mantle control of Mars and Earth, it is common for the horizontal variation of the CMB heat flux to be described in terms of spherical harmonics. Especially for Mars, the majority of studies rely on spherical harmonics of degree $l=1$ and order $m=0$, i.e., a simple cosine variation (e.g., Stanley et al.\ 2008\nocite{Stanley2008}, Amit et al.\ 2011\nocite{Amit2011}, Aurnou and Aubert 2011\nocite{Aurnou2011}, DW13). \rev{Notable exceptions} are the study by Sreenivasan and Jellinek (2012)\nocite{Sreenivasan2012}, in which a localised temperature anomaly was used, \rev{and that by Monteux et al.\ (2015), in which the anomaly pattern was derived from realistic impact models}. \rev{The former} study relies on fixed temperature boundary conditions, basal heating, and strong CMB temperature anomalies and is thus quite different from our systematic approach, \rev{which features fixed flux boundary conditions, internal heating, and anomaly amplitudes not exceeding the mean CMB heat flux}. Here, we explore more locally confined variations of the CMB heat flux. The thermal CMB anomaly $q^\prime$ is characterised by four parameters: its amplitude $q^\ast$, its opening angle $\Psi$, and the colatitude and longitude ($\tau, \phi_0$) of its midpoint. The anomaly has the form
\begin{equation}
 q^\prime = \begin{cases} \frac{1}{2} \left( \cos\left(2\pi\frac{\alpha}{\Psi}\right) + 1 \right) \mbox{ for  } \alpha <\Psi \\ 
0 \mbox{ else}\, \end{cases}  
\end{equation}
where $\alpha(\theta,\phi)$ is the opening angle between the central vector $\vec r(\theta,\phi)$ and the mid-point vector $\vec r (\tau, \phi_0)$. Furthermore, the anomaly is normalised, such that the net heat flux through the CMB is constant. The total CMB heat flux is then given by the mean heat flux $q_0$, the anomaly $q^\prime$, and the normalisation \rev{$C(\Psi)$ as
\begin{equation}
q_{cmb} = q_0 - q^\ast \left( \frac{q^\prime}{C(\Psi)} -  1 \right) \ ,
\end{equation} 
with
\begin{equation}
 C(\Psi) = \int_{S_\Psi} q^\prime \sin \theta d \theta d \phi \ = \frac{1-\cos \Psi}{4} \ ,
\end{equation}
where $S_\Psi$ is the area of the anomaly up to its rim, which is given by $\Psi$.}
 Note that the case of ($\Psi=180^\circ$, $\tau=0$) is identical to the spherical harmonic $Y_{10}$ mentioned above. Figure \ref{figqprof} shows the CMB heat flux profiles for $\tau=0$, $q^\ast=1$, and various widths $\Psi$. For such parameters, the heat flux anomaly is axisymmetric and reduces the heat flux at the northern pole to zero.

\subsection*{Numerical model and runs}
The magnetohydrodynamics (MHD) equations (Eqs. \ref{nseq}, \ref{heat}, and \ref{ind_eq}) were numerically solved using the MagIC3 code in its shared memory version (Wicht 2002\nocite{wicht2002}, Christensen and Wicht 2007\nocite{Christensen2007b}). The numerical resolution is given by 49, 288, and 144 grid points in the radial direction, the azimuthal direction, and along the latitude, respectively. For the higher Rayleigh number cases, the numerical resolution was increased to 61, 320, and 160 points, respectively. 

We conducted a broad parameter study in which eight anomaly widths between $\Psi=180^\circ$ (planetary scale) and $\Psi=10^\circ$ (most concentrated hot spot) were used. Furthermore, we tested four different anomaly amplitudes ranging from $q^\ast=0.2$ to $q^\ast=1$. The peak position of the anomaly was tilted at six different angles between $\tau=0$ (polar anomaly) and $\tau=90^\circ$ (equatorial anomaly). \rev{Because the ancient Martian core was fully liquid at the time the magnetisation was acquired, the thick shell regime is investigated here. Particularly, an inner core with an Earth-like aspect ratio of $\beta=0.35$ is retained in most of the models to ensure consistency with the existing literature regarding Earth and Mars. However, the influence of the aspect ratio $\beta$ was tested by varying the aspect ratio between $\beta=0.35$ and $\beta=0.15$, which corresponds to the smallest inner core size.} The vigour of the convection was varied using four different Rayleigh numbers from $\mathrm{Ra}=2 \times 10^7$, which is only slightly supercritical, to $\mathrm{Ra}=1.6 \times 10^8$, which ensures rich turbulent dynamics. We repeated the numerical experiments for simulations including the magnetic field for various amplitudes $q^\ast$ and tilt angles $\tau$. Together with reference cases with no \rev{thermal} boundary \rev{heterogeneity} for each $\beta$ and $\mathrm{Ra}$, these parameters represent 129 hydrodynamic and 73 \rev{self-consistent dynamo} simulations (202 simulations in total).

Each case was time integrated until a statistically steady state was reached and then time averaged over a significant fraction of the viscous (magnetic) diffusion time. The moderate Ekman number allowed each of the hydro cases to be modelled within a computational time of one or two days when parallelised over 12 cores, whereas \rev{dynamo} simulations usually require several days.

\rev{ \section*{Previous Work and Output Parameters} }
\begin{table}[!ht]
\caption{Flow symmetries. Time-averaged relative kinetic energy of various flow contributions (C) obeying different symmetries, which are represented by three-letter abbreviations. The first two letters indicate the equatorial symmetry: equatorial symmetric (ES) or equatorial antisymmetric (EA). The last letter indicates the axisymmetry: axisymmetric (A) or non-axisymmetric (N). The first two rows are magnetic runs (Figure \ref{figmp}), and the other four are hydro runs (Figure \ref{figflowpol}). Parameters: $\mathrm{Ra}= 4 \times 10^7$.}
      \begin{tabular}{cccccc}
        \hline
        $q^\ast$ &$\Psi$     & $C_{\mathrm{ESA}}$  & $C_{\mathrm{ESN}}$  & $C_{\mathrm{EAA}}$ & $C_{\mathrm{EAN}}$\\ \hline
        0        &0          & 0.03 & 0.72 & 0.01 & 0.24 \\
        1        &$180^\circ$&0.03 & 0.09 & 0.78 & 0.10\\ \hline
        0.5      &$180^\circ$& 0.05 & 0.29 & 0.34 & 0.31 \\
        0.75     &$120^\circ$& 0.05 & 0.31 & 0.33 & 0.30 \\
        1        &$90^\circ$ & 0.05 & 0.35 & 0.31 & 0.24 \\
        1        &$30^\circ$ & 0.04 & 0.61 & 0.08 & 0.27 \\ \hline
\end{tabular}
\label{tabsym}
\end{table} 

  \begin{figure}[!ht]
\includegraphics[width=0.9\textwidth]{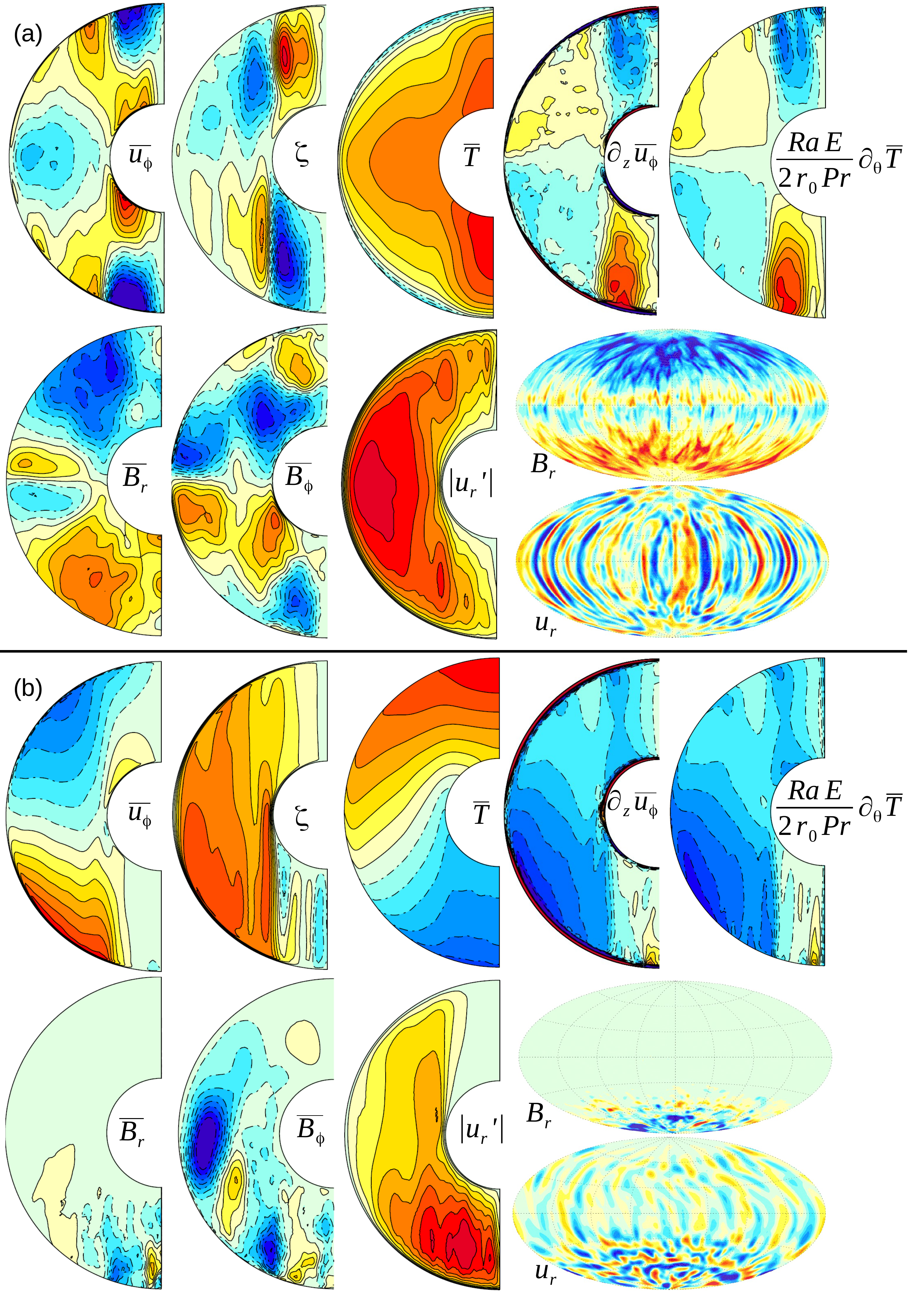}
  \caption{\csentence{Mean flow and magnetic properties}. (a) Homogeneous reference case ($q^\ast=0$, $\mathrm{Ra}=4 \times 10^7$). (b) Standard boundary forcing case ($q^\ast=1.0$, $\tau=0$, $\Psi=180^\circ$). The first row in each part shows, from left to right, the zonal flow ($\overline{u_\phi}$), the stream function of the meridional circulation ($\zeta$), the zonal temperature ($\overline{T}$) and the two sides of the thermal wind equation ($\partial \overline{u_\phi} / \partial z$ and $\mathrm{Ra E}/(2 r_0 \mathrm{Pr}) \, \partial T / \partial \theta$). The second row in each part contains the radial field ($\overline{B_r}$), the mean azimuthal field ($\overline{B_\phi}$), the intensity of non-axisymmetric radial flows ($|u_r^\prime|$), and hammer projections of the radial field at the surface (top) and the radial flow at mid-depth (bottom). \rev{Parameters: $\mathrm{Ra}=4 \times 10^7$, $q^\ast=1.0$, $\tau=0$, $\beta=0.35$, $\mathrm{Pm}=2$.} }
\label{figmp}
\end{figure}

Figure \ref{figmp} illustrates the mean flow and field properties for an unperturbed reference dynamo case with homogeneous boundary heat flux (\ref{figmp},a) and a commonly studied model with a heterogeneous CMB heat flux (\ref{figmp},b). In the reference case, ESN convective columns (e.g., Busse 1970\nocite{Busse1970}) account for 72\% of the total kinetic energy. Table \ref{tabsym} shows the kinetic energy symmetry contributions in the two cases. Axisymmetric flow contributions consist of the zonal flow and the meridional circulation. Both are predominantly equatorially symmetric but have low amplitudes. These equatorially symmetric, axisymmetric (ESA) kinetic energy contributions therefore amount to only 3\% of the total kinetic energy (Table \ref{tabsym}). The zonal temperature $\overline{T}$ is also equatorially symmetric, and its colatitudinal gradient is in very good agreement with the $z$-variation of the zonal flow (last two plots in the first row of Figure \ref{figmp},a), which proves that this $z$-variation is caused by thermal wind. The mean radial and azimuthal magnetic fields (first two plots in the second row of Figure \ref{figmp}) show the typical equatorial antisymmetry of a dipole-dominated magnetic field. This field is produced by non-axisymmetric flows (intensity $| u_r^\prime|$) in an $\alpha^2$-mechanism (Olson et al.\ 1999\nocite{Olson1999}). The final two plots in the second row of Figure \ref{figmp} show the Hammer--Aitoff projections of the radial field at the CMB and the radial flow at mid-depth.

We compare the homogeneous case with the most commonly studied CMB heat flux anomaly: a $Y_{10}$ anomaly with a strong amplitude ($q^\ast=1.0$). Such an anomaly cools the southern hemisphere more efficiently than the northern hemisphere; hence, the temperature decreases from the hot north to the cool south, leading to a negative temperature gradient along the colatitude. Such strong temperature anomalies are known to modify the leading order vorticity balance between the pressure and the Coriolis force. The curl of the Navier--Stokes equation (\ref{nseq}) gives to first order (neglecting viscous and inertia terms): 
\begin{equation}
 \nabla \times \hat z \times \vec u = \frac{\mathrm{Ra E}}{2 \mathrm{Pr}} \frac{1}{r_{cmb}} \nabla \times (\vec r T) \ .
\label{eqtw1}
\end{equation}
In models with a homogeneous heat flux and relatively small Ekman number, the right-hand side of eq. \ref{eqtw1} is small, and the flow (at least the convective flow) tends to be vertically invariant. Because of the rigid walls applied here, the zonal flow is weak and \rev{a}geostrophic in the reference case. However, for the boundary-forced system, the right-hand side becomes large. The heterogeneous CMB heat flux mainly cools the southern hemisphere and leaves the northern hemisphere hot. The large-scale temperature asymmetry that develops between the north and south is responsible for driving a significant axisymmetric thermal wind. For the mean azimuthal component of eq. \ref{eqtw1}, it is thus found that 
\begin{equation}
 \frac{\partial \overline{u_\phi}}{\partial z} = \frac{\mathrm{Ra E}}{2 \mathrm{Pr}} \frac {1}{r_{cmb}} \frac{\partial \overline{T}}{\partial \theta} \ .
\label{eqtw}
\end{equation}
Figure \ref{figmp} compares the right- and left-hand sides of eq. \ref{eqtw} and demonstrates that this thermal wind balance is indeed well fulfilled in both the homogeneous and $Y_{10}$ cases. The larger north--south gradient in the latter case drives a zonal wind system with retrograde and prograde jets in the northern and southern hemispheres, respectively. This EAA flow system dominates the kinetic energy once the amplitude of the boundary pattern is sufficiently large (see Table \ref{tabsym}). 
We therefore quantify the influence of the boundary forcing by measuring the relative contribution $C_{\mathrm{EAA}}$ of EAA flows by evaluating (in spectral space) the relative kinetic energy of spherical harmonic flow contributions of order $m=0$ (axisymmetric) and odd degree $l=2n+1$ (equatorially antisymmetric)
\begin{equation}
 C_{\mathrm{EAA}}=  \frac{ \int_{r_{icb}}^{r_{cmb}} \sum_l E_{2l+1,0}^k r^2 dr }{ \int_{r_{icb}}^{r_{cmb}} \sum_{l,m} E_{l,m}^k r^2 dr}  \ .
\end{equation}
The radial flows required to induce radial fields are then concentrated in the southern polar region (Figure \ref{figmp}, bottom row), where the cooling is more efficient. Hence, the induced radial field is also concentrated in the respective hemisphere. The dominant magnetic field production, however, remains the thermal wind shear, which produces a strong axisymmetric azimuthal field via the $\Omega$-effect. The bottom row of Figure \ref{figmp} shows the hemisphere-constrained radial field and the amplified azimuthal toroidal field created by shearing motions around the equator. DW13 reported that the $\Omega$-effect dominates once the $Y_{10}$ amplitude exceeds 60\% of the mean heat flux. The dynamo then switches from $\alpha^2$- to $\alpha\Omega$-type, initiating periodic polarity reversals that are characteristic of this dynamo type (see also Dietrich et al.\ 2013\nocite{Dietrich2013b}). Note that EAA symmetric flows can emerge in dynamo models with a homogeneous heat flux (Landeau and Aubert 2011\nocite{Landeau2011})\rev{ when they are internally heated and satisfy a large convective supercriticality.} Conversely, the anomalies themselves can also drive sufficiently complex flows to induce magnetic fields in the absence of \rev{thermal and compositional buoyancy fluxes} (Aurnou and Aubert 2011\nocite{Aurnou2011}).

\rev{The magnetic field is stronger in the convectively more active hemisphere and shows weaker magnetic flux in the less active hemisphere. Amit et al.\ (2011) gave an estimate of the mean crustal magnetisation per hemisphere on present-day Mars, which can be compared to the geometry of the model output fields. DW13 defined the magnetic field hemisphericity $H_{sur}$ at the planetary surface as:
\begin{equation}
 H_{sur} = \left\vert \frac{ B_N - B_S}{B_N+B_S} \right\vert \ ,
\label{eqdefH}
\end{equation}
where $B_N$ and $B_S$ are the radial magnetic fluxes in the northern and southern hemispheres, respectively. Note that the crustal value is $0.55 \pm 0.1$ (Amit et al.\ 2011, DW13). As radial fields are induced by convective flows, an equivalent convective hemisphericity $\Gamma$ can be defined as
\begin{equation}
 \Gamma = \left\vert \frac{ \Gamma_N - \Gamma_S}{\Gamma_N+\Gamma_S} \right\vert \ .
\end{equation}
A formal derivation of $\Gamma_N$ and $\Gamma_S$ is given in the Results section. We further quantify the mean flow amplitude with the hydrodynamic Reynolds number Re for the hydrodynamic simulations and the magnetic Reynolds number Rm for the dynamo simulations. In the latter case, the mean core magnetic field strength is given by the Elsasser number $\Lambda$:
\begin{equation}
 \mathrm{Re}  = \frac{UD}{\nu} \ , \qquad \mathrm{Rm} = \frac{UD}{\lambda} \ ,\qquad \Lambda = \frac{B^2}{\mu_0 \lambda \rho \Omega} 
\end{equation}
}

As suggested in the study by Hori et al.\ (2014) and DW13, heat flux anomalies applied along the equator ($\tau=90^\circ$) yield special solutions. In this case, the equatorial symmetry of the CMB heat flux is re-established as in the case with homogeneous boundaries, but the azimuthal \rev{symmetry} and axisymmetry are broken. It has been reported that anomalies of a planetary scale modelled by spherical harmonics of degree and order unity ($Y_{11}$) lead to flows dominated in spectral space by azimuthal order $m=1$ resembling the anomaly. In the same manner as Hori et al.\ (2014), we measure the dominance of $m=1$ by evaluating
\begin{equation}
 E_{m1} =  \frac{\int_{r_{icb}}^{r_{cmb}} \sum_l E_{l,1}^k r^2 dr}{\int_{r_{icb}}^{r_{cmb}}\sum_{l,m} E_{l,m}^k r^2 dr}  \ .
\label{eqdefem1}
\end{equation}

\section*{Results}
\rev{Of the 202 simulations performed in this study, select numerical models are presented in Table \ref{taball} to provide an overview of and compare a number of hydrodynamic runs (centre column) and their equivalent dynamo runs (right column). All results in Table \ref{taball} were calculated with fixed parameters $\mathrm{Pr}=1$, $\mathrm{Pm} = 0/2$ (hydrodynamic/dynamo runs), $\mathrm{Ra} = 4 \times 10^7$, $\mathrm{E} = 10^{-4}$, and $\beta=0.35$ and with variable forcing amplitude $q^\ast$, anomaly width $\Psi$, and tilt angle $\tau$.}
\begin{table}[!ht]
\caption{Select numerical models with fixed Rayleigh $\mathrm{Ra}=4 \times 10^7$, Ekman $\mathrm{E}=10^{-4}$, and Prandtl $\mathrm{Pr}=1$ numbers and aspect ratio $\beta=0.35$. The magnetic Prandtl number was kept constant at $\mathrm{Pm}=2$ throughout all magnetic simulations. The quantities of entries marked with a ``*'' were not calculated. This applies to both convective $\Gamma$ and magnetic $H_{sur}$ hemisphericity. If the magnetic field reverses, the oscillation frequency is given in multiples of $2\pi \mathrm{Pm}/\tau_{vis}$. Note that for $q^\ast=1$, $\Psi=180^\circ$, and $\tau=45^\circ$, the field reverses but with an unclear frequency.} 
      \begin{tabular}{ccc|cccc|ccccccc} 
                  &        &         & hydrodynamic &     &     &          & magnetic &     &     &          &           &            & \\ 
        \hline
         $q^\ast$ & $\Psi$ [$^\circ$] & $\tau$ [$^\circ$] & Re    & $C_{\mathrm{EAA}}$ & $E_{m1}$ & $\Gamma$ & Rm       & $C_{\mathrm{EAA}}$ & $E_{m1}$ & $\Gamma$ & $\Lambda$ & $H_{sur}$ & $\omega$ \\ 
        \hline
         0        &  -     &   -     & 154.7 & 0.066 & 0.029 & 0.057 & 224.3 & 0.007 & 0.064 & 0.010 & 7.83 & 0.005 & -\\ 
         1        &  180   &   0     & 223.3 & 0.603 & 0.013 & 0.686 & 446.7 & 0.747 & 0.015 & 0.778 & 1.71 & 0.573 & 49.42  \\ 
                  &        &   22    & 220.9 & 0.589 & 0.019 & *     & 433.5 & 0.738 & 0.022 & *     & 2.50 & 0.561 & 46.54 \\ 
                  &        &   45    & 213.3 & 0.535 & 0.040 & *     & 406.3 & 0.691 & 0.037 & *     & 3.28 & 0.431 & ?  \\ 
                  &        &   63    & 198.7 & 0.375 & 0.081 & *     & 347.4 & 0.594 & 0.071 & *     & 7.94 & 0.220 & -\\ 
                  &        &   90    & 181.0 & 0.029 & 0.239 & *     & 308.7 & 0.003 & 0.257 & *     & 2.72 & *     & -\\ 
                  &  120   &   0     & 208.0 & 0.427 & 0.023 & 0.432 & 428.0 & 0.759 & 0.002 & 0.684 & 3.94 & 0.415 & 25.58 \\ 
                  &        &   22    & 208.9 & 0.419 & 0.031 & *     & 412.4 & 0.729 & 0.024 & *     & 4.12 & 0.410 & 22.74 \\ 
                  &        &   45    & 198.1 & 0.370 & 0.071 & *     & 360.1 & 0.654 & 0.046 & *     & 7.32 & 0.226 & -\\ 
                  &        &   63    & 184.9 & 0.229 & 0.145 & *     & 300.1 & 0.386 & 0.097 & *     & 8.68 & 0.136 & -\\
                  &        &   90    & 168.8 & 0.016 & 0.247 & *     & 255.6 & 0.006 & 0.257 & *     & 7.84 & *     & -\\ 
                  &  90    &   0     & 192.5 & 0.291 & 0.029 & 0.265 & 365.3 & 0.703 & 0.003 & 0.501 & 7.94 & 0.318 & -\\ 
                  &        &   22    & 191.4 & 0.293 & 0.038 & *     & 369.3 & 0.698 & 0.028 & *     & 3.69 & 0.297 & -\\ 
                  &        &   45    & 184.0 & 0.243 & 0.102 & *     & 291.9 & 0.483 & 0.055 & *     & 12.87& 0.170 & -\\ 
                  &        &   63    & 177.4 & 0.149 & 0.164 & *     & 283.9 & 0.262 & 0.111 & *     & 6.98 & 0.118 & -\\
                  &        &   90    & 163.2 & 0.013 & 0.231 & *     & 239.7 & 0.006 & 0.238 & *     & 2.40 & *     & -\\ 
                  &  60    &   0     & 172.1 & 0.169 & 0.032 & 0.194 & 271.8 & 0.224 & 0.144 & 0.194 & 10.19& 0.077 & -\\ 
                  &        &   22    & 172.0 & 0.178 & 0.357 & *     & 263.5 & 0.258 & 0.111 & *     & 11.8 & 0.065 & -\\ 
                  &        &   45    & 170.4 & 0.157 & 0.093 & *     & 249.1 & 0.205 & 0.075 & *     & 11.97& 0.060 & -\\ 
                  &        &   63    & 168.9 & 0.109 & 0.151 & *     & 249.5 & 0.109 & 0.155 & *     & 8.66 & 0.047 & -\\
                  &        &   90    & 158.0 & 0.015 & 0.175 & *     & 230.3 & 0.006 & 0.237 & *     & 0.99 & *     & -\\ 
                  &  30    &   0     & 158.5 & 0.083 & 0.031 & 0.065 & 232.5 & 0.010 & 0.047 & 0.057 & 7.12 & 0.007 & -\\ 
                  &        &   22    & 158.7 & 0.091 & 0.027 & *     & 233.6 & 0.009 & 0.031 & *     & 6.70 & 0.007 & -\\ 
                  &        &   45    & 158.1 & 0.093 & 0.031 & *     & 239.1 & 0.017 & 0.130 & *     & 6.97 & 0.012 & -\\ 
                  &        &   63    & 157.5 & 0.087 & 0.045 & *     & 238.2 & 0.019 & 0.167 & *     & 7.54 & 0.011 & -\\
                  &        &   90    & 151.9 & 0.027 & 0.062 & *     & 228.7 & 0.006 & 0.199 & *     & 8.55 & *     & -\\ 
        \hline
         0.75     &  180   &   0     & 208.4 & 0.473 & 0.015 & 0.532 & 425.0 & 0.763 & 0.019 & 0.767 & 3.69 & 0.413 & 34.91\\ 
                  &  120   &   0     & 198.1 & 0.340 & 0.025 & 0.377 & 368.8 & 0.716 & 0.022 & 0.462 & 7.80 & 0.327 & -\\ 
                  &  90    &   0     & 184.5 & 0.250 & 0.028 & 0.245 & 346.5 & 0.666 & 0.023 & 0.401 & 7.11 & 0.284 & -\\ 
                  &  60    &   0     & 168.1 & 0.154 & 0.030 & 0.163 & 255.2 & 0.136 & 0.132 & 0.043 & 9.45 & 0.054 & -\\ 
                  &  30    &   0     & 158.0 & 0.085 & 0.027 & 0.065 & 231.2 & 0.008 & 0.053 & 0.010 & 7.09 & 0.005 & -\\ 

        \hline
         0.5      &  180   &   0     & 193.9 & 0.351 & 0.023 & 0.383 & 347.3 & 0.725 & 0.031 & 0.555 & 10.75& 0.258 & -\\ 
                  &  120   &   0     & 181.8 & 0.257 & 0.028 & 0.272 & 307.2 & 0.609 & 0.033 & 0.379 & 14.44& 0.193 & -\\ 
                  &  90    &   0     & 173.7 & 0.195 & 0.027 & 0.187 & 269.9 & 0.321 & 0.071 & 0.193 & 12.36& 0.110 & -\\ 
                  &  60    &   0     & 163.5 & 0.128 & 0.027 & 0.115 & 245.9 & 0.078 & 0.116 & 0.052 & 8.72 & 0.035 & -\\ 
                  &  30    &   0     & 157.3 & 0.078 & 0.027 & 0.083 & 227.7 & 0.009 & 0.064 & 0.038 & 7.81 & 0.007 & -\\ 
        \hline
         0.25     &  180   &   0     & 166.0 & 0.173 & 0.028 & 0.206 & 239.5 & 0.183 & 0.086 & 0.2   & 13.20 & 0.060 & -\\ 
                  &  120   &   0     & 163.5 & 0.147 & 0.027 & 0.163 & 237.8 & 0.110 & 0.117 & 0.111 & 11.08 & 0.052 & -\\ 
                  &  90    &   0     & 161.6 & 0.122 & 0.028 & 0.095 & 237.7 & 0.074 & 0.105 & 0.048 & 9.32 & 0.033 & -\\ 
                  &  60    &   0     & 158.2 & 0.089 & 0.026 & 0.070 & 231.7 & 0.009 & 0.092 & 0.070 & 7.59 & 0.015 & -\\ 
                  &  30    &   0     & 155.3 & 0.067 & 0.026 & 0.043 & 227.7 & 0.009 & 0.068 & 0.024 & 7.58 & 0.008 & -\\ 
        \hline

\end{tabular}
\label{taball}
\end{table}

\subsection*{Amplitude}
As noted above, the amplitude of the CMB heat flux variation is determined by the thermal lower mantle structure. For Earth, the amplitude can exceed the super-adiabatic part of the homogeneous flux, indicating that values of $q^\ast>1$ may be possible. Here, we restricted ourselves to $q^\ast\leq 1$, thereby avoiding models with local core heating that may lead to stable stratification.

\begin{figure}[!ht]
\includegraphics[width=0.8\textwidth]{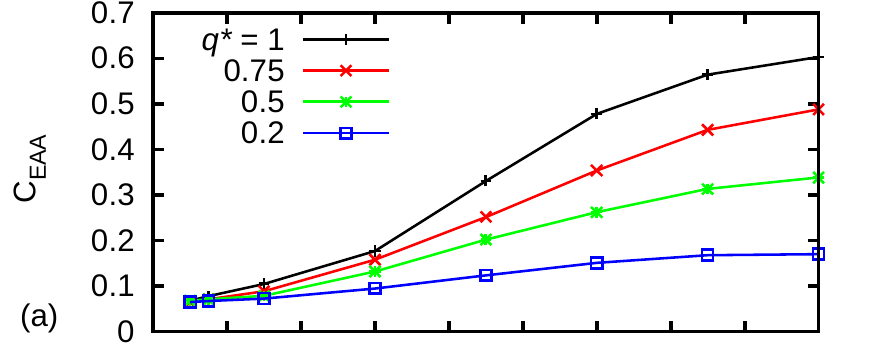}
\includegraphics[width=0.8\textwidth]{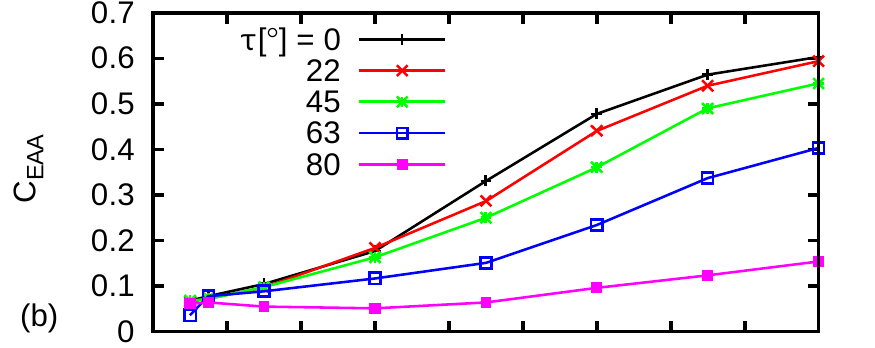}
\includegraphics[width=0.8\textwidth]{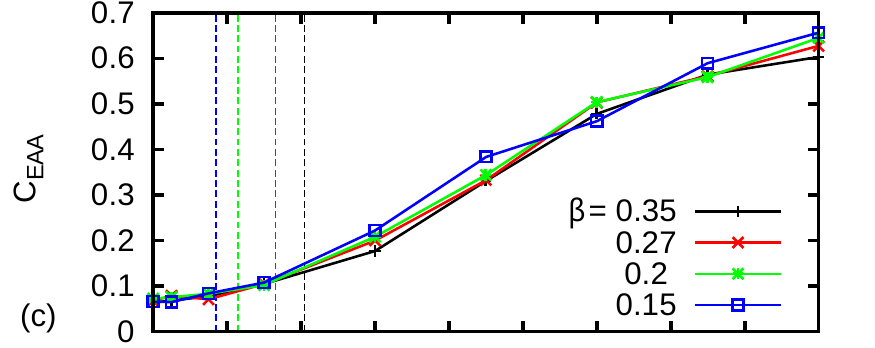}
\includegraphics[width=0.8\textwidth]{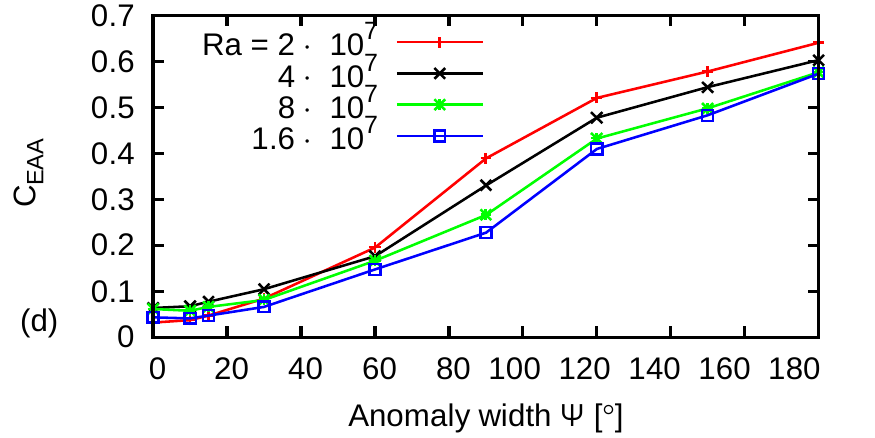}
  \caption{\csentence{Influence of various model parameters on EAA contribution}. Impact of (a) anomaly amplitude, (b) tilt angle, c) shell geometry, and d) vigour of convection on the EAA contribution for varying anomaly width $\Psi$. The coloured vertical lines denote the colatitude of the tangent cylinder. \rev{Reference parameters (unless specified otherwise): $\mathrm{Ra}=4 \times 10^7$, $q^\ast=1.0$, $\tau=0$, $\beta=0.35$.}}
\label{figpareaa}
\end{figure}

To isolate the influence of the anomaly amplitude, the tilt angle was fixed at $\tau=0$, and the amplitude $q^\ast$ and the width $\Psi$ were varied in small steps. Figure \ref{figpareaa}(a) shows the strength of the EAA contribution $C_{\mathrm{EAA}}$ with respect to the total kinetic energy. For the largest and strongest anomaly with $\Psi=180^\circ$ and $q^\ast=1$, the highest value of $C_{\mathrm{EAA}}$ is found (black line in Figure \ref{figpareaa}(a)). When the width $\Psi$ of the anomaly was reduced and its amplitude was kept fixed, the EAA symmetry contribution reduced accordingly. Hence, the enormous EAA contributions found in DW13 are strongly related to the large scale of the anomaly chosen there. It is thus not (or not only) the breaking of the equatorial symmetry that leads to strong antisymmetric flow contributions. An anomaly with a weaker amplitude ($q^\ast < 1$) reduces the strength of the EAA contribution $C_{\mathrm{EAA}}$ almost linearly. Interestingly, halving the anomaly width has almost the same effect as halving the anomaly amplitude. As an example, for $\Psi=180^\circ$ and $q^\ast=0.5$, the EAA contribution is 0.338, whereas $\Psi=90^\circ$ and $q^\ast=1.0$ yield an EAA contribution of 0.331. 

\begin{figure}[!ht]
\includegraphics[width=0.8\textwidth]{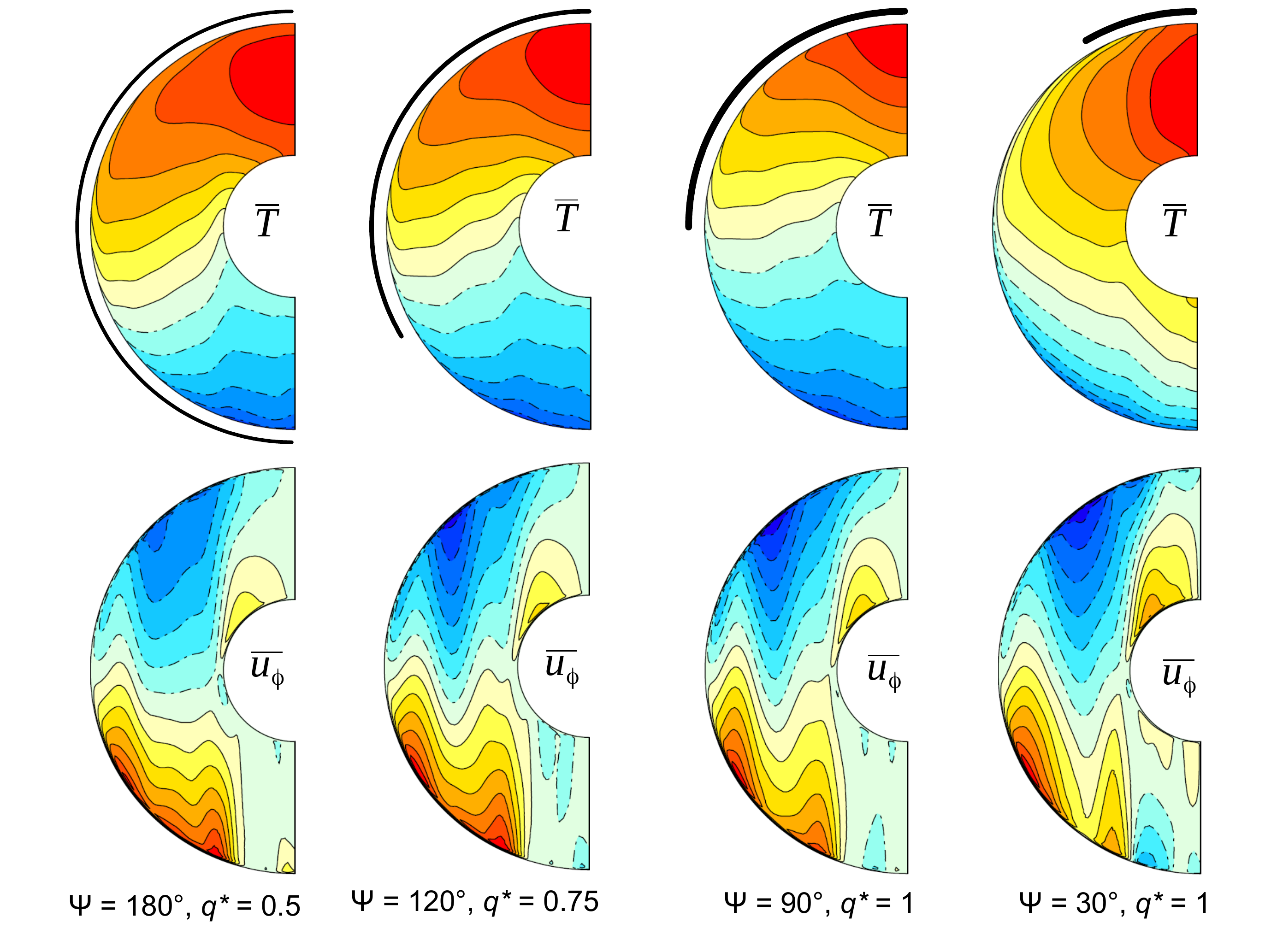}
  \caption{\csentence{Zonal temperature and flow}. The zonally averaged temperature (top) and differential rotation (bottom) are shown for four example cases with variable anomaly width $\Psi$ and amplitude $q^\ast$. Note that the first three cases have similar EAA symmetric flows. The black arcs denote the width of the anomaly, and the arc thickness \rev{scales with} the amplitude. \rev{Parameters: $\mathrm{Ra}=4\times 10^7$, $\tau=0$.} }
	\label{figflowpol}
\end{figure}

Figure \ref{figflowpol} shows the zonally averaged temperatures and azimuthal flows for various combinations of anomaly amplitudes $q^\ast$ and widths $\Psi$ in four selected models with polar anomalies ($\tau=0$). Note that the thermal wind balance (eq. \ref{eqtw}) is always satisfied. One could expect that narrower anomalies (smaller $\Psi$) with stronger horizontal heat flux gradients would have stronger and more concentrated thermal winds, but this does not seem to be the case. The large-scale temperature anomaly between the north and south develops independent of the anomaly width. The first three models in Figure \ref{figflowpol} have similar EAA contribution strengths $C_{\mathrm{EAA}}$ (see also Table \ref{tabsym}), supporting the quasi-linear increase of EAA symmetry with amplitude and width.

\subsection*{Latitudinal Position}
Thus far, we have focused on polar anomalies, where the \rev{anomaly peak vector} is aligned with the \rev{rotation axis}, which is a special situation. Because those break only the equatorial symmetry and not the azimuthal symmetry, the total CMB heat flux remains colatitude-dependent but axisymmetric. Mantle plumes and giant impacts generally do not sit on or reach the pole. Therefore, we explored several tilt angles $\tau$. The magnetic case with a planetary-scale heat flux anomaly was explored in DW13 across a variety of tilt angles, and it was demonstrated that all tilt angles $\tau < 80^\circ$ lead to a fairly strong EAA contribution. Such behaviour was also observed in a study by Amit et al.\ (2011).

Figure \ref{figpareaa}(b) shows the EAA flow contribution $C_{\mathrm{EAA}}$ for four tilt angles $\tau$. The first case, with $\tau=22^\circ$, was set up such that the anomaly peak vector was located at the colatitude of the tangent cylinder. As will be discussed later, the tangent cylinder and the shell geometry were expected to have a strong influence on the dynamics, but our results show that they have no particular influence on the location of the heat flux anomaly \rev{for the thick shells studied here}. However, tilting the anomaly further from the axis of rotation did not significantly affect the strength of the EAA symmetry for moderate tilt angles ($\tau \leq 45^\circ$). In this case, non-axisymmetric flow contributions are changed little, and the EAA contribution remains surprisingly strong. For $\Psi=180^\circ$, we can decompose the heat flux anomaly into $Y_{10}$ and $Y_{11}$ contributions. The effective $Y_{10}$ contribution is simply $q^\ast \cos(\tau)$. The EAA remained significantly stronger than even the effective $Y_{10}$ contribution would suggest. For example, for $\tau=45^\circ$ we expected $C_{\mathrm{EAA}} = C_{\mathrm{EAA}}(\tau=0)/ \sqrt{2} = 0.42$, but we obtained $C_{\mathrm{EAA}}=0.54$. The nearly equatorial case showed distinct behaviour. In general, it might be concluded that any anomaly smaller than $\Psi = 60^\circ$  might only weakly influence the core convection. Note this behaviour is independent of the tilt angle or anomaly amplitude.  
  \begin{figure}[!ht]
\includegraphics[width=0.8\textwidth]{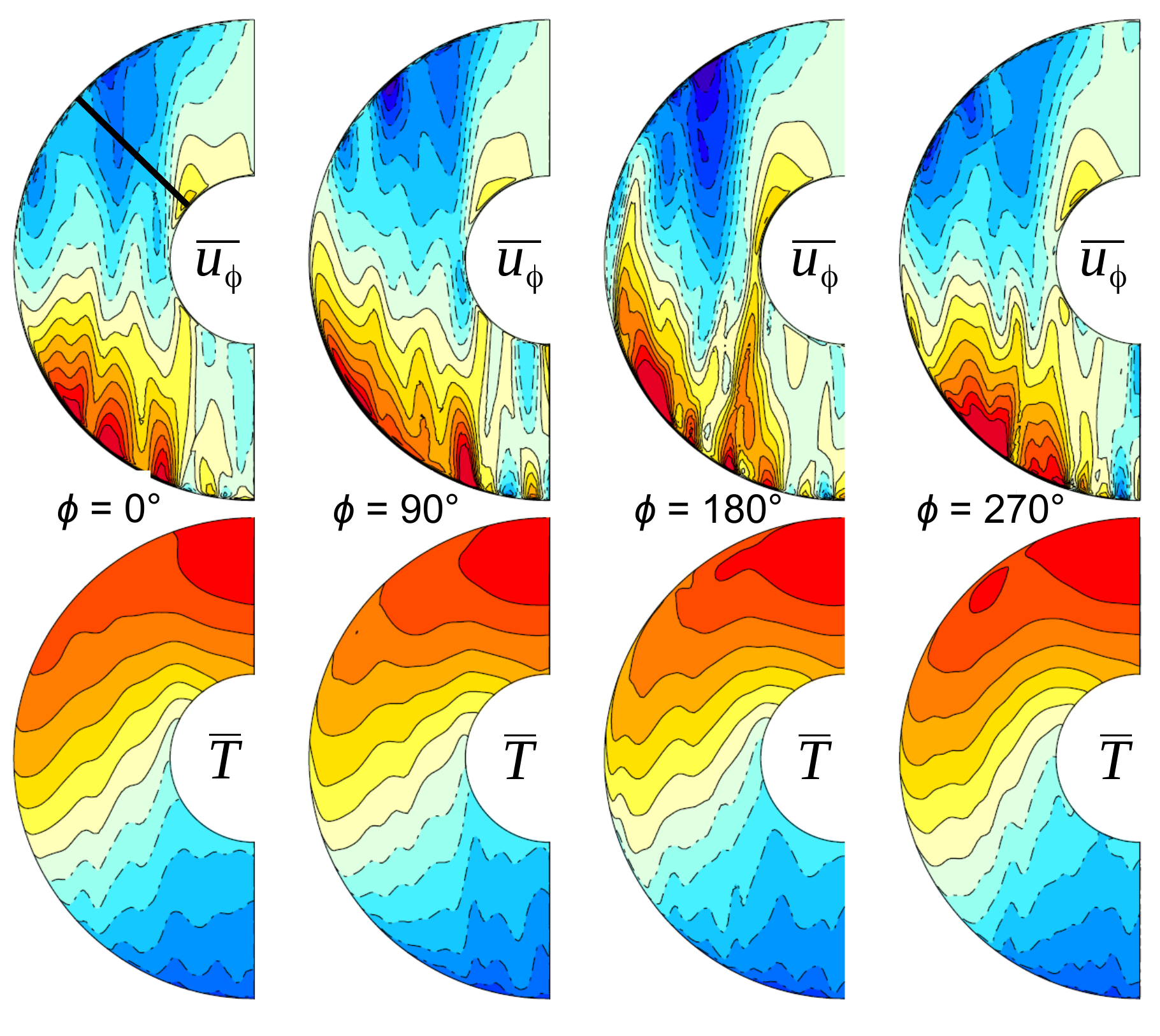}
  \caption{\csentence{Meridional cuts of flow and temperature}. The temperature (top) and differential rotation (bottom) are shown for four different meridional cuts at $\phi$ for an anomaly of width $\Psi=180^\circ$, amplitude $q^\ast=1$, and tilt angle $\tau=45^\circ$. The thick black line denotes the central anomaly peak vector. \rev{Parameters: $\mathrm{Ra}=4\times 10^7$.}}
\label{figflowtilt}
\end{figure}

Figure \ref{figflowtilt} illustrates the time-averaged temperature and zonal flow in four different meridional cuts. The sample model here is for $q^\ast=1$, $\Psi=180^\circ$, and $\tau=45^\circ$. The first plot is positioned at a longitude of $\phi=0$ and includes the location of the centre of the anomaly (black line). At this value of $\phi$, the CMB heat flux at $\theta=45^\circ$ is exactly zero. The three other plots are taken at intervals of $\phi=90^\circ$ eastwards. Remarkably, large-scale equatorial asymmetry in the temperature is visible in all cuts; hence, the EAA symmetry driven by the thermal wind is a meaningful measure for the tilted cases as well. Interestingly, not much is visible of the broken azimuthal symmetry. 

\subsection*{Aspect Ratio}
For numerical reasons, we have kept in our model an inner core that is purely driven by internal heat sources. Hori et al.\ (2010)\nocite{Hori2010} have shown that such an inner core has little impact on the solution for homogeneous outer boundary conditions. Figure \ref{figpareaa}(c) proves that this remains true for the inhomogeneous heat flux explored here. The aspect ratio affects the critical Rayleigh number for the onset of convection. Therefore, the cases compared here have different super-criticality. However, this also seems to have little impact for the limited range of Rayleigh numbers we studied. 

\subsection*{Vigour of Convection}
The Rayleigh number Ra can influence the EAA instability in at least two ways. It directly scales buoyancy forces and thus also scales the thermal wind strength (see eq. \ref{eqtw}). The EAA contribution should therefore grow linearly with Ra. However, Ra also controls the convective vigour and length scale. As Ra grows, the scale decreases while the vigour increases, and both of these changes lead to more effective mixing, which should counteract the impact of the inhomogeneous boundary condition. Figure \ref{figpareaa}(d) illustrates that the EAA contribution decreases with growing Ra such that the latter effect seems to dominate. However, for large anomalies, the relative EAA flow symmetry seems to become saturated for all Ra.

\subsection*{Magnetic Fields}
\label{secgamma}
It has been reported that EAA flows enforced by boundary anomalies dramatically affect the morphology and time dependence of the magnetic field \rev{(DW13)}. A polar planetary-scale anomaly with amplitude $q^\ast=1$ and width $\Psi=180^\circ$ transforms a strong and stationary dipolar-dominated magnetic field into a weaker wave-like hemispherical dynamo that is dominated by an axisymmetric toroidal field (DW13). The altered magnetic field configuration also indicates the strength of the EAA symmetry. DW13 suggested that the magnetic field significantly enhances the EAA contribution. A flow that inhibits a strong EAA symmetry results from a strongly asymmetric temperature anomaly\rev{, which is maintained by the thermal wind} and hence confines the convective motions into a single hemisphere. This weakens the global amount of convection, potentially inducing magnetic fields. Another possible effect is that the magnetic field may relax the strong rotational constraint on the flow, thus allowing the convection to develop in more three-dimensional rather than columnar structures. This would support the convection in the magnetically active hemisphere and increase the temperature asymmetry and hence the EAA symmetry. Finally, the strong azimuthal toroidal field that emerges in the equatorial region, created by fierce equatorial antisymmetric shear associated with the EAA symmetric differential rotation, can potentially suppress columnar convective flows within this equatorial region (DW13). 

  \begin{figure}[!ht]
\includegraphics[width=0.9\textwidth]{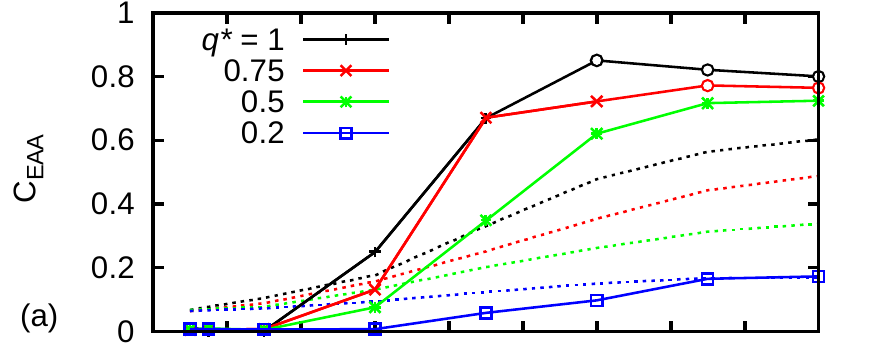}
\includegraphics[width=0.9\textwidth]{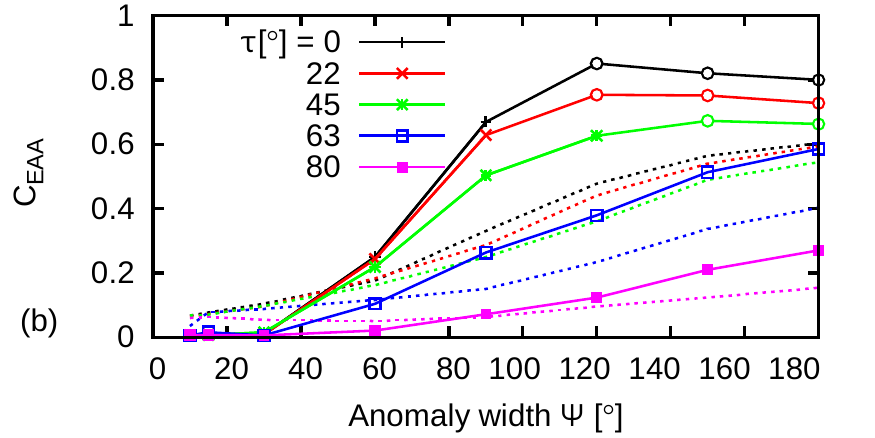}
  \caption{\csentence{Effect of magnetic fields on EAA convective mode}. Relative EAA kinetic energy contribution for (a) polar anomalies (fixed $\tau=0$) with various amplitudes $q^\ast$ and for (b) fixed $q^\ast=1.0$ with various tilt angles $\tau$ as function of the anomaly width $\Psi$. For comparison, the hydrodynamic reference cases are included as dashed curves. The EAA contributions $C_{\mathrm{EAA}}$ for the homogeneously cooled reference cases are 0.01 (dynamo) and 0.06 (hydrodynamic). Oscillatory dynamos are indicated by empty circles. \rev{Parameters: $\mathrm{Ra}=4\times 10^7$, $\mathrm{Pm}=2$ (dynamo) or $\mathrm{Pm}=0$ (hydrodynamic).}}
\label{figmampp}
\end{figure}

Figure \ref{figmampp} compares simulations including the magnetic field generation process (solid lines) with the previously discussed hydrodynamic cases (dashed). Figure \ref{figmampp}(a) is restricted to polar anomalies (fixed $\tau=0$) and takes several anomaly amplitudes $q^\ast$, whereas Figure \ref{figmampp}(b) fixes the amplitude $q^\ast$ and varies the tilt angle $\tau$; both parts of the figure plot the EAA contribution against the anomaly width $\Psi$. For smaller or vanishing anomalies, the magnetic field actually suppresses the effect of the heat flux inhomogeneity and reduces the EAA contribution to nearly zero. The magnetic field now further suppresses the equatorially asymmetric contributions that were already relatively weak in the non-magnetic case. 
  \begin{figure}[!ht]
\includegraphics[width=0.31\textwidth]{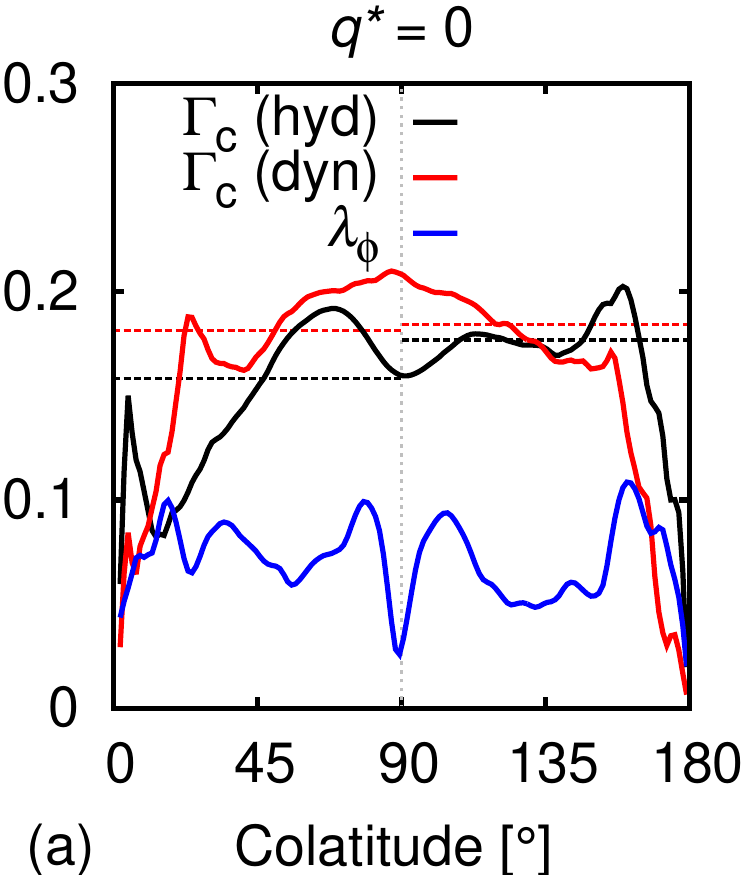}
\includegraphics[width=0.31\textwidth]{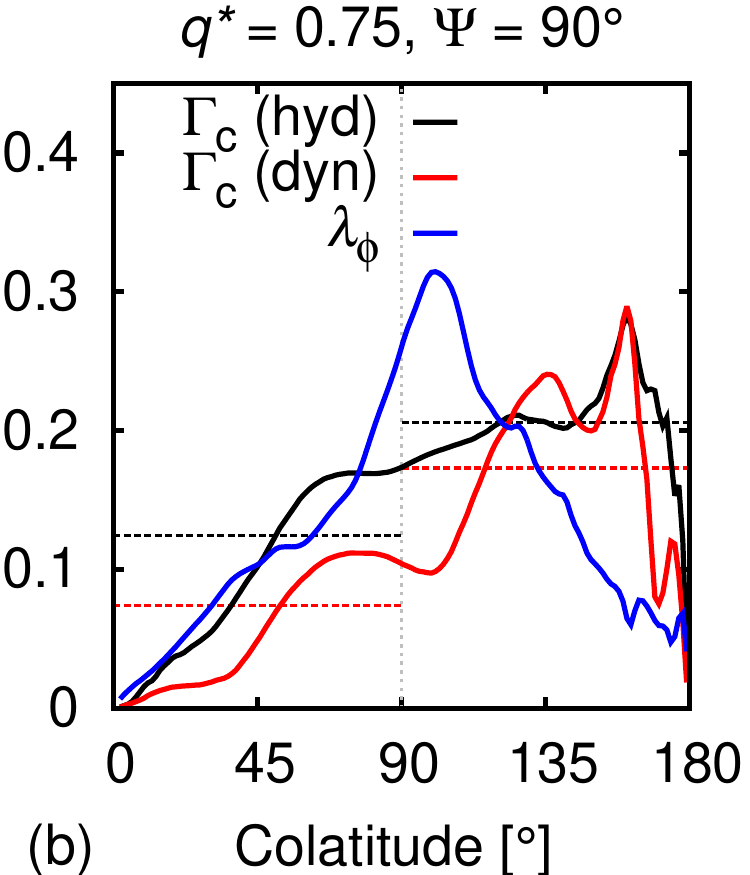}
\includegraphics[width=0.31\textwidth]{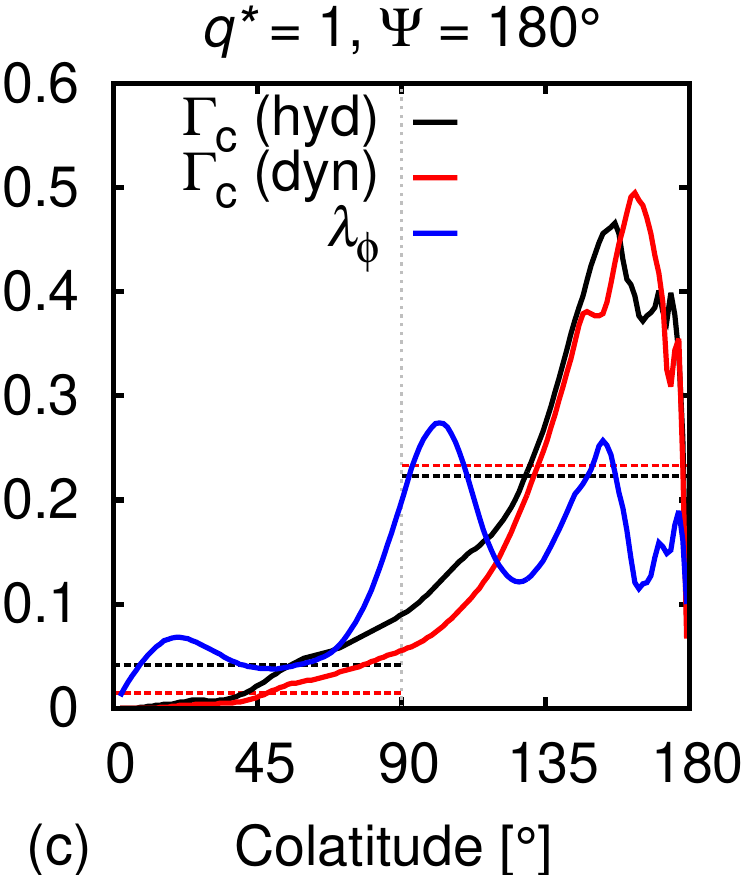}
  \caption{\csentence{Radially integrated convective heat transport}. Colatitudinal profiles of the vertical convective heat transport $\Gamma_c$ for hydrodynamic (black) and magnetic (red) simulations. The horizontal dashed lines denote the hemispherical average of each $\Gamma_c$. Blue lines denote the relative contribution $\lambda_\phi$ of the mean azimuthal toroidal field. (a) Reference case with homogeneous boundary heat flux. (b) $\Psi=90^\circ$ and $q^\ast=0.75$. (c) $Y_{10}$ case with a large-scale anomaly: $\Psi=180^\circ$ and $q^\ast=1.0$. Note that whereas (a) and (b) show long time averages, (c) is calculated from a few snapshots because of the reversing magnetic field. \rev{Parameters: $\mathrm{Ra}=4\times 10^7$, $\tau=0$, $\mathrm{Pm}=2$.}}
\label{figgampa}
\end{figure}

  \begin{figure}[!ht]
\includegraphics[width=0.9\textwidth]{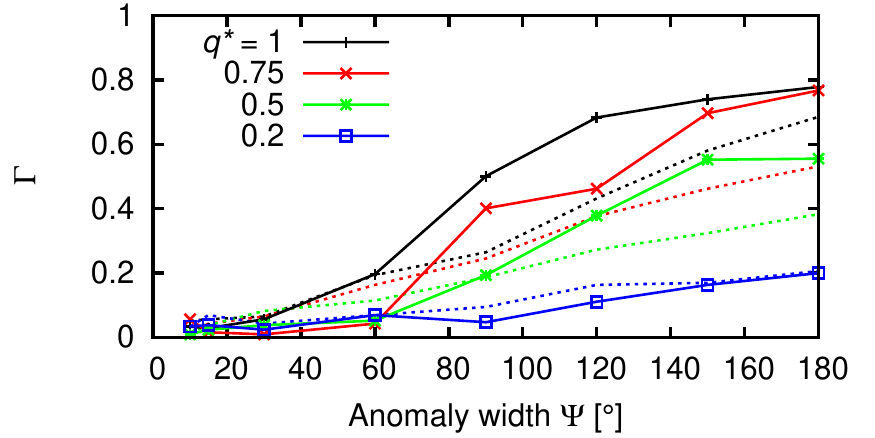}
  \caption{\csentence{Hemisphericity of vertical heat transport}. Asymmetry of the vertical heat transport in the northern and southern hemispheres. \rev{Pure hydrodynamic models (dashed) and dynamo simulations (solid) are included. Parameters: $\mathrm{Ra}=4\times 10^7$, $\tau=0$, $\mathrm{Pm}=2$ (dynamo) or $\mathrm{Pm}=0$ (hydrodynamic).}}
\label{figgamma}
\end{figure}

If the anomalies reach a width of $\Psi=90^\circ$, the magnetic field drastically enhances the EAA symmetry relative to the hydrodynamic runs. For $q^\ast=0.75$ and $\Psi=90^\circ$, the magnetic field changed the EAA contribution from 0.25 to 0.67, indicating a stronger convection in the magnetically more active (southern) hemisphere. All dynamos reach a strong magnetic field (Elsasser number \rev{$\Lambda \ge 1$}), which is when the leading force balance in the momentum equation is between the Coriolis and Lorentz forces. Note that the azimuthal vorticity balance (thermal wind) is still exclusively between the Coriolis force and the buoyancy (see the last two plots in the upper row of the lower set in Figure \ref{figmp}).

Figure \ref{figmampp}(b) illustrates that the magnetic effect on the EAA contribution remains similar as the tilt angle $\tau$ varies. The magnetic field increases the EAA contribution for larger opening angles but decreases it for smaller values of $\Psi$. Open circles in Figure \ref{figmampp} show the locations at which oscillatory reversing dynamos have been found. This behaviour is promoted by a strong $\Omega$-effect and therefore requires a large thermal wind shear. Because the main thermal wind contribution is EAA, the respective measure of the EAA symmetry is a good proxy for the $\Omega$-effect (Dietrich et al.\ 2013). Figure \ref{figmampp} demonstrates that oscillatory dynamos are only found for relatively large EAA contributions. The tilt angle also plays a role in this behaviour. For $\tau=0$, only models with $C_{\mathrm{EAA}}>0.7$ are oscillatory, whereas for $\tau=45^\circ$, a non-reversing dynamo still exists at $C_{\mathrm{EAA}}=0.68$. Even though a strong EAA symmetric flow leads to strong shear around the equatorial region, yielding a strong $\Omega$-effect (Dietrich et al.\ 2013\nocite{Dietrich2013b}), it can co-exist with non-reversing fields. If, despite the strong shearing, a sufficient fraction of toroidal field is created by non-axisymmetric helical flows, the field remains stable.

For a more detailed investigation of how the magnetic field affects the flow, a measure of the heat transport efficiency was developed. We correlate the convective motion in terms of non-axisymmetric radial flows $u_r^\prime$ with non-axisymmetric temperature perturbations $T^\prime$ over the azimuth and time. Radial integration of this measure gives the mean vertical convective heat transport $\Gamma_c$ as a function of the colatitude alone:
\begin{equation}
 \Gamma_c (\theta) = \int_{r_i}^{r_o} \overline{ u_r^\prime T^\prime}(r,\theta) \, r^2 \, dr \ ,
\end{equation}
where the overbar denotes the correlation over the azimuthal angle $\phi$ and time. This measure is closely related to the definition of the Nusselt number proposed by Otero et al.\ (2002)\nocite{Otero2002}, where another integration along the colatitude is taken. Because of the large temperature variation along the colatitude, we keep $\Gamma_c$ as a function of the colatitude $\theta$. Figure \ref{figgampa} shows the colatitudinal profiles of $\Gamma_c$ for a few cases; in each panel, the hydrodynamic (black) and self-consistent dynamo (red) simulations are shown. Of all the simulations, we chose to investigate the reference case with homogeneous heat flux, the case with $\Psi=90^\circ$ and $q^\ast=0.75$, and the most commonly studied case ($Y_{10}$) given by $\Psi=180^\circ$ and $q^\ast=1.0$ (Figure \ref{figgampa}(a), (b), and (c), respectively). The study case in Figure \ref{figgampa}(b) was selected because it shows an enormous difference between the hydrodynamic and dynamo simulations and the magnetic field is non-reversing. The case in Figure \ref{figgampa}(c) features a reversing magnetic field and hence oscillates between weak (hydrodynamic) and strong (dynamo) EAA symmetry. The correlation needed to calculate $\Gamma_c$ in Figure \ref{figgampa}(c) is taken over only three time steps, during which the magnetic field remains strong and does not change sign, whereas for the other cases, the correlations are taken over the full magnetic diffusion time with tens of snapshots. 

DW13 suggested that the emerging strong \rev{axisymmetric} toroidal field \rev{induced by the EAA shear} suppresses the radial, \rev{non-axisymmetric} convective flows. To test this hypothesis, we calculated the strength of the mean azimuthal magnetic field relative to the total magnetic field. Because some magnetic fields oscillate, we took the time-averaged \rev{root mean square} azimuthal field rather than the time-averaged field and obtained:
\begin{equation}
 \lambda_\phi(\theta) = \frac{1}{\sqrt{B^2}} \int_{r_i}^{r_o} \sqrt{ \overline{B_\phi}^2}(r,\theta) \, r^2 \, dr \ ,
\end{equation}
which is included in Figure \ref{figgampa} in blue. As suggested in DW13, we found that $\lambda_\phi$ is large when the reduction of convective flows is large (see Figure \ref{figgamma}). The vertical convective heat transport $\Gamma_c$ is clearly suppressed when the toroidal field indicated by $\lambda_\phi$ is large. For the case in Figure \ref{figgampa}(c), the enhancement of $\Gamma_c$ in the magnetically more active hemisphere is visible as well.

Furthermore, we defined the convective hemisphericity $\Gamma$, \rev{which is equivalent to the magnetic field hemisphericity $H_{sur}$ but is based on the vertical heat transport} $\Gamma_c$ integrated over either the northern or southern hemisphere, such that \rev{
\begin{equation}
 \Gamma = \left\vert \frac{ \Gamma_N - \Gamma_S}{\Gamma_N+\Gamma_S} \right\vert \ \mbox{\, , \, where \, \, \, } \Gamma_c^{N,S} = \int_{0,\pi/2}^{\pi/2,\pi} \Gamma_c(\theta) \sin \theta d \theta \ .
\end{equation}}
These values are plotted in Figure \ref{figgampa} as dashed vertical lines, and the \rev{convective hemisphericity} $\Gamma$ is plotted in Figure \ref{figgamma} for all polar anomalies ($\tau=0$). According to our results, it can be concluded that the magnetic field (mainly \rev{the axisymmetric part of} $B_\phi$) is responsible for the increased equatorial asymmetry of the convective cooling and thus the boost of EAA symmetric flows \rev{(see Figure \ref{figmampp})}. For the tilted cases (Figure \ref{figmampp}(b)), the magnetically driven EAA enhancement also appeared for all tilt angles. Even for $\tau=63^\circ$, the EAA contribution reached nearly 0.6 for the largest anomaly width $\Psi$. However, for these tilted cases, the EAA mode increased linearly with $\Psi$, where for smaller tilt angles, saturation occurred. 

\subsection*{Equatorial anomalies}
\begin{figure}[!ht]
\includegraphics[width=0.9\textwidth]{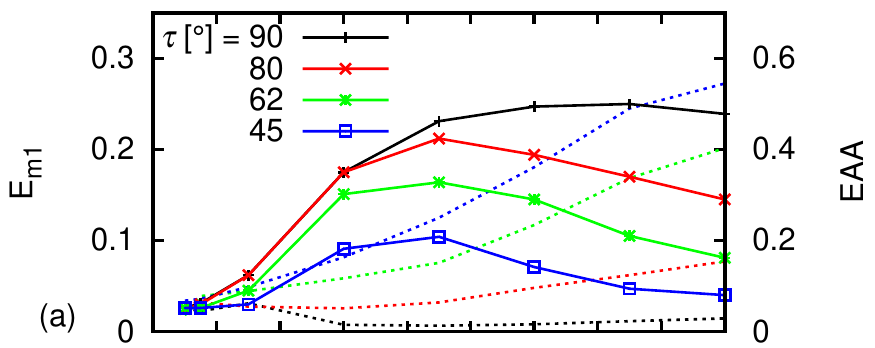}
\includegraphics[width=0.9\textwidth]{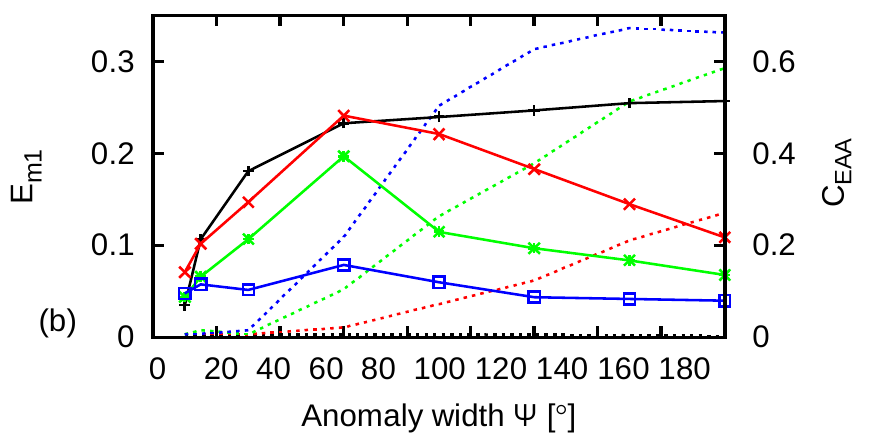}
  \caption{\csentence{$m=1$ dominance for equatorial anomalies}. Relative kinetic energy contribution of the $m=1$ flows ($E_{m1}$, solid lines) and the EAA contribution (dashed) for anomalies of various tilt angles ($\tau$) as a function of anomaly width $\Psi$ for (a) hydrodynamic and (b) self-consistent dynamo simulations. \rev{Parameters: $\mathrm{Ra}=4\times 10^7$, $q^\ast=1.0$, $\mathrm{Pm}=2$ (dynamo) or $\mathrm{Pm}=0$ (hydrodynamic).}}
\label{figmone}
\end{figure}

\rev{For anomalies tilted towards the equator ($\tau=90^\circ$), the flows show a strong azimuthal alteration along with the outer boundary heat flux. Hence, significant kinetic energy contributions are expected from flows with an equivalent spectral order of $m=1$ (see definition of $E_{m1}$, eq. \ref{eqdefem1})}. Figure \ref{figmone} gives an overview of the \rev{spectral response} $E_{m1}$ for the hydrodynamic cases with an anomaly amplitude of $q^\ast=1$. Large equatorial anomalies led to an increase from $E_{m1}=0.026$ in the homogeneous reference case to $ E_{m1}=0.25$ for the largest anomaly width. If the anomaly is tilted away from the equator again, both the equatorial and azimuthal symmetry are broken. Hence, it would be expected that apart from the equatorial case with $\tau=90^\circ$, smaller tilt angles will also show an enhancement of $E_{m1}$. Figure \ref{figmone} also shows various tilt angles between $\tau=90^\circ$ and $45^\circ$, where the $E_{m1}$ amplitude decreases with decreasing tilt angle. Interestingly, the cases with more concentrated anomalies, e.g., $\Psi=90^\circ$, yielded higher values of $E_{m1}$ than the planetary-scale anomalies with $\Psi=180^\circ$. For larger $\Psi$ reaching far enough across the equator, it seems that the equatorial asymmetry takes control, and EAA symmetric flows are established. It then seems reasonable for $E_{m1}$ to decay only for $\tau \neq 90^\circ$. If the magnetic field is included (Figure \ref{figmone}(b)), the systematic behaviour of $E_{m1}$ found in the hydrodynamic simulations is rather equivalent. This suggests that the magnetic field is not important for $E_{m1}$ as a measure of the dynamic response to breaking the azimuthal symmetry of the outer boundary heat flux. 

\subsection*{Application to Mars}
For application to Mars, the time-averaged surface hemisphericity of the radial field is correlated with the EAA symmetry. The two are dynamically linked because a strong EAA kinetic energy contribution relies on strongly equatorially asymmetric convection, which in turn induces a hemispherical field. The magnetic field is extrapolated by a potential field towards the Martian surface with an outer core radius of $r_{cmb}=1680\, \mathrm{km}$ and a surface radius of $r_{sur}=3385\, \mathrm{km}$. The magnetic field hemisphericity $H_{sur}$ at the surface gives a ratio that is a function of the radial field intensities $B_N$ and $B_S$ integrated over the northern and southern hemispheres, as defined in eq. \ref{eqdefH}. 
\begin{figure}[!ht]
\includegraphics[width=0.9\textwidth]{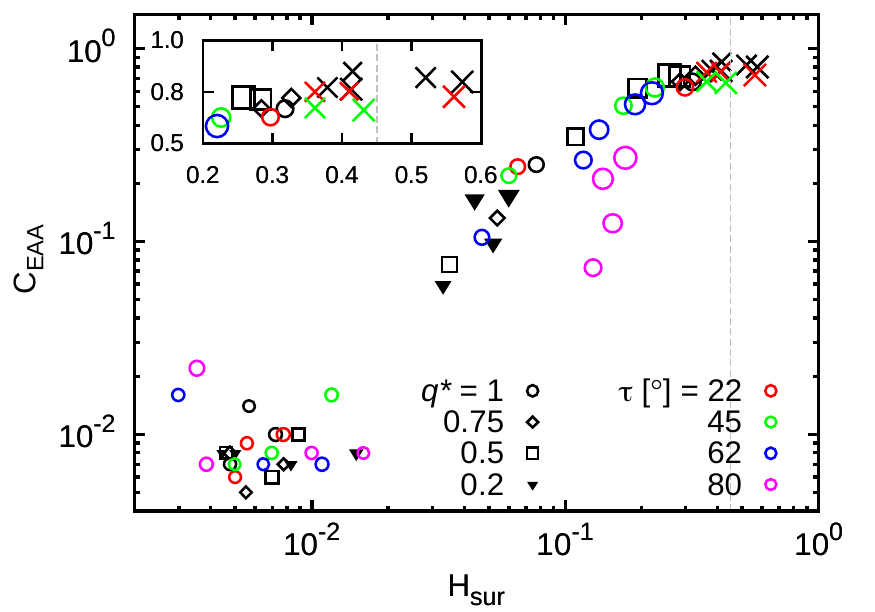}
  \caption{\csentence{Magnetic field hemisphericity vs.\ EAA}. Hemisphericity of the radial magnetic field at the Martian surface plotted against the EAA flow contribution. The different colours, symbols, and symbol sizes represent different tilt angles ($\tau$), anomaly amplitudes ($q^\ast$), and widths of the anomaly ($\Psi$), respectively and the crosses indicate reversing dynamos. \rev{The specific values of $C_\mathrm{EAA}$ and further characteristic quantities for the oscillatory dynamos can be found in Figure \ref{figmampp} and Table \ref{taball}.} The small inset plot (top left) includes only points with $H_{sur} \geq 0.2$ and $C_{\mathrm{EAA}} \geq 0.5$. The vertical grey dashed line indicates the lower limit of the Martian crustal value of $H_{sur}$. \rev{Parameters: The Rayleigh and Ekman numbers are kept constant throughout all simulations in the plot ($\mathrm{Ra}=4\times 10^7$, $\mathrm{E}=10^{-4}$)}.}
\label{fighsur}
\end{figure}

Figure \ref{fighsur} shows the correlation between the EAA symmetry and the surface hemisphericity of the radial field. In these simulations, a weak EAA led to weak $H_{sur}$, as \rev{expected}. For large EAA symmetry enforced by boundary forcing, the magnetic fields tend to be more hemispherical. The lower bound on the Martian crustal value of $H_{sur}$ is 0.45 (Amit et al.\ 2011), which is only reached when the anomalies have a large horizontal extent ($\Psi > 120^\circ$) and amplitude ($q^\ast  \geq 0.75$). Even though the EAA contribution can become dominant for smaller and weaker anomalies, it is far more challenging to induce a magnetic field of sufficient surface hemisphericity in this case. Furthermore, all magnetic fields with sufficiently high $H_{sur}$ values are oscillatory, which makes a thick unidirectional magnetisation unlikely (see discussion in DW13). 

\revb{DW13 explored the parameter dependence of a polar $Y_{10}$ anomaly, changing the Rayleigh, magnetic Prandtl, and Ekman numbers within the numerically accessible limits (see their Figure 11). Not unexpectedly, decreasing the Ekman number led to smaller hemisphericities because the geostrophic geometry was more severely enforced. This can be counteracted by increasing the anomaly amplitude; however, increasing Ra or Pm also seemed to help, likely because inertia or Lorentz forces more significantly contribute to balancing the Coriolis force. At realistically small Ekman numbers of approximately $3\times10^{-15}$ and appropriate Ra and Pm for Mars, this likely means that unrealistically large heat flux variations would be required to yield the observed hemisphericity. Inertial forces are thought to be small in planetary cores, whereas Lorentz forces should not significantly exceed the strength reached at the smaller Ekman number of $10^{5}$ explored by DW13. 

The generally oscillatory nature of high hemisphericities remains a problem at all parameter combinations and geometries explored in DW13 and in the present study. Latitudinal temperature variations paired with their respective gradients in convective efficiency never fail to drive strong thermal winds. These in turn lead to a significant $\Omega$-effect that seems to favour oscillatory dynamos (Dietrich et al.\ 2013). None of the variations in the general set-up explored in this paper have indicated that this fundamental mechanism is incorrect.}

\section*{Conclusions}
We constructed a suite of 202 numerical models of spherical shell convection and magnetic field generation in which the outer boundary heat flux was perturbed by an anomaly of variable width, amplitude, and position. \rev{The convection was driven exclusively by secular cooling, which is an appropriate model for terrestrial planets in the early stages of their evolution, when no inner core is present.} The dynamic response of the flow was measured in terms of the expected spectral equivalent of the boundary forcing. For anomalies breaking the equatorial symmetry, the relative strength of EAA kinetic energy was used, and for anomalies breaking the azimuthal symmetry, the relative strength of flows with a spectral order of $m=1$ was investigated ($E_{m1}$). 

\rev{For hydrodynamic models without a magnetic field, the strength of the EAA symmetry was found to increase almost linearly with the amplitude and width of the anomaly. These flows are driven by a large-scale equatorial asymmetry in the axisymmetric temperature field. Hence, a more localised CMB heat flux anomaly does not lead to a stronger or more confined thermal wind, even though the horizontal variation of the heat flux is locally much larger. The simulations also indicated that models perturbed by narrower anomalies or anomalies that are not aligned with the axis of rotation also yield the same fundamental temperature asymmetry. For example, if the anomaly peak is tilted by an angle $\tau \leq 45^\circ$ from the axis of rotation, the EAA symmetry is quite similar to that in the case of the polar anomaly ($\tau=0$). Furthermore, this suggests that the system is more sensitive to changes in the equatorial symmetry than in the azimuthal symmetry.}

For equatorial anomalies ($\tau=90^\circ$), the spectral response ($E_{m1}$) reached up to 25\% of the kinetic energy and was only mildly affected by the magnetic field. Interestingly, for tilt angles $45^\circ \leq \tau \leq 80^\circ$, the contribution of $E_{m1}$ could be measured as well and was found to typically be the strongest for moderately sized anomalies $60^\circ \leq \Psi \leq 90^\circ$. Larger anomalies broke the equatorial symmetry as well, thus increasing the EAA symmetry \rev{at the cost of the $E_{m1}$ symmetry}. 

\rev{For numerical reasons, the models were run with an inner core present; as such, we further tested the influence of smaller aspect ratios, which proved to be negligible. A similar conclusion was reached by Hori et al.\ (2010) for a homogeneous outer boundary heat flux and may be extended with this study to boundary-forced models. As the primary purpose of our model is to comprehensively parametrise the various boundary anomalies, out of various other model parameters (Ra, E, $\mathrm{Pm}$, Pr), we only tested the influence of increasing the convective vigour. Our model also indicated that stronger convective stirring does not suppress fundamental temperature asymmetries or the EAA mode. However, it was shown that the model is slightly more sensitive to boundary forcing when convective driving is weaker. DW13 provides a discussion of the possible dependence on the Ekman number (rotation rate) and the magnetic Prandtl number, showing the robustness of a similar model featuring $Y_{10}$-forcing.}

\rev{In the presence of dynamo action}, the behaviour is far more nonlinear. For narrow anomalies, the flow is equatorially symmetrised by the dipole field, whereas anomalies with widths $\Psi \geq 90^\circ$, amplitudes $q^\ast \geq 0.5$, and tilt angles $\tau \leq 45^\circ$ strongly boost the EAA symmetry. It was shown that the strong azimuthal toroidal field around the equator suppresses the remaining columnar convection and further increases the asymmetry \rev{in the temperature and, as a consequence, the antisymmetry in the core flow}. \rev{Hence,} the magnetic field prevents narrow heat flux anomalies from affecting the core convection, where it drastically increases their respective effects when they reach a horizontal extent on the same order as the core radius. This effect is independent of the tilt angle, amplitude, and width of the anomaly when $\tau \leq 45^\circ$, $q^\ast \geq 0.5$, and $\Psi \geq 90^\circ$. For all models within these boundaries, this in turn also implies that CMB heat flux anomalies can be smaller, weaker, and non-polar while still yielding effects similar to those of the fundamental $Y_{10}$ anomaly. A similar observation was reported recently by Monteux et al.\ (2015). Hence, our results suggest that the core dynamos of \rev{ancient} Mars or early Earth are sensitive to CMB heat flux anomalies only if they are strong in amplitude and large in horizontal extent. 

Regarding the hemispherical magnetisation of the Martian crust, it seems rather unlikely that the magnetising field is as hemispherical as the crustal pattern suggests; \revb{hence, the crustal magnetisation dichotomy seems only realistically explained by additional demagnetisation events of external origins in the northern hemisphere.} \revb{The results of the numerical models clearly indicate that} a sufficiently hemispherical field is possible only if the anomaly is of core scale and significantly affects the CMB heat flux. However, all of these geometrically corresponding hemispherical dynamos show rather frequent polarity \rev{reversals} and hence would require a crustal rock magnetisation time on the order of the magnetic diffusion time (tens of thousands of years), \rev{which} might \rev{be} a rather unrealistic scenario \rev{for a thick magnetised layer of at least 20\,km}. \rev{Note that it is possible to create a magnetic field that shows a smaller degree of equatorial asymmetry {\it and} is stable in time.} \revb{However, at more realistic model parameters, e.g., a smaller Ekman number, it seems likely that these models remain applicable only when a much stronger forcing is applied to sufficiently break the typical $z$-invariance of the flow (geostrophy). One common feature consistently found here and in DW13 is that a stronger thermal forcing naturally leads to oscillatory fields.}

Thus, the main results obtained in this study are:\rev{
\begin{itemize}
 \item Fundamental large-scale equatorial asymmetry in the temperature and hence EAA symmetric flows emerge independent of the width, position, and amplitude of a CMB heat flux anomaly. 
 \item The magnetic field prevents narrow heat flux anomalies from affecting the core convection but drastically increases their respective effects when the anomalies reach horizontal extents on the same order as the core radius.
 \item \revb{At least for the parameters and geometries explored here and in DW13, it is not possible for a hemispherical dynamo to explain the observed dichotomy in Martian crustal magnetisation.} 
\end{itemize}
}


\begin{backmatter}

\section*{Authors' contributions}

WD proposed the topic, designed the study and carried out the 
numerical experiments. JW developed the numerical implementation and extended the code to include the variable anomaly. KH and JW collaborated with WD in the construction of the manuscript. All authors read and approved the final manuscript.

\section*{Acknowledgements}

This work was supported in part by the Science and Technology Facilities Council (STFC). Numerical simulations were undertaken on ARC1, part of the High Performance Computing facilities at the University of Leeds, UK.


\bibliographystyle{peps-art} 
\bibliography{sedi}  

\end{backmatter}
\end{document}